\documentclass[10pt,letterpaper]{article}
\usepackage{graphicx}\usepackage{epstopdf} 
\usepackage{xcolor}
\usepackage{opex3}
\usepackage{cite}
\usepackage{bm}
\usepackage{color} 
\usepackage{framed} 
\usepackage{comment} 
	\definecolor{shadecolor}{gray}{0.875}
	\excludecomment{comment}  

\newcommand{\e}{{\rm e}}
\renewcommand{\d}{{\rm d}}
\newcommand{\x}{{\bf x}}

\newcommand{\dydxv}[2]{ {{\partial #1} \over {\partial #2}} }

\begin{document}

\title{Parity-Time Symmetric Coupled Microresonators with a Dispersive Gain/Loss}

\author{Sendy Phang$^{1,*}$, Ana Vukovic$^1$, Stephen Creagh$^2$, Trevor M. Benson$^1$, Phillip D. Sewell$^1$, Gabriele Gradoni$^2$}

\address{$^1$George Green Institute for Electromagnetics Research, Faculty of Engineering, \\University of Nottingham,Nottingham, NG7 2RD, UK\\
$^2$School of Mathematical Sciences, University of Nottingham,\\Nottingham. NG7 2RD, UK}

\email{$^*$sendy.phang@nottingham.ac.uk} 


\begin{abstract}
The paper reports on the coupling of Parity-Time (PT)-symmetric whispering gallery resonators with realistic material and gain/loss models. Response of the PT system is analyzed for the case of low and high material and gain dispersion, and also for two practical scenarios when the pump frequency is not aligned with the resonant frequency of the desired whispering gallery mode and when there is imbalance in the gain/loss profile. The results show that the presence of dispersion and frequency misalignment causes skewness in frequency bifurcation and significant reduction of the PT breaking point, respectively. Finally, as coupled WGM resonators are inherently lossy structures, we show that unbalancing the gain/loss in resonators  is required to compensate for inherent loss of the structure and achieve improved PT properties. 
\end{abstract}


\section{Introduction}
Photonics is emerging as a popular practical platform for the exploration of Parity-Time (PT)-symmetric systems characterized by balanced loss and gain and having a threshold point at which real eigenfrequencies of the system coalesce to become complex conjugates \cite{bender99,lin11,chong11,benisty12}. The existence of this threshold point is essential to the unique properties of PT-structures, such as unidirectional invisibility and simultaneous lasing and absorption\cite{lin11,chong11}. This phenomena opens new avenues for the realization of practical devices such as lasers, optical memory, optical switches and logic-gates\cite{lupu13,nazari11,hodaei14,longhi14,phang13,phang15,kulishov13}. To date, PT-symmetric structures based on Bragg gratings and coupled optical systems have been investigated both theoretically \cite{lin11,chong11,phang14a,benisty12,jones12,ctyroky10,ctyroky14,ganainy07,greenberg04,longhi10,longhi14,nolting96} and experimentally \cite{ruter10,peng14,peng14b,regensburger13,chang14,feng14}. Recently, a PT-symmetric system based on two coupled microresonators and two fiber-taper waveguides has been experimentally demonstrated and shown to exhibit direction-dependent behavior at a record low power of 1$\mu$W \cite{peng14,feng14}. This is primarily attributed to strong field localization and build up of energy in the resonant whispering gallery modes\cite{peng14,feng14}, and has further strengthened the argument for using resonant structures rather than waveguides as building blocks of PT-symmetric systems. In contrast to PT symmetric coupled waveguide systems where the eigenmodes are purely real below the threshold point, the PT symmetric coupled microresonators have complex eigenfrequencies below the threshold point due to inherent radiation losses\cite{chong11,benisty12,longhi10,ctyroky10}. 

In this paper we investigate the fundamental properties of the PT resonant system based on two coupled whispering gallery resonators within the context of both realistic material properties and practical operating constraints. In particular we discuss how practical dispersive properties of material gain and loss that satisfy the Kramers-Kronig relationship affect the performance of microcavity-based PT resonant structures. Our surprising conclusion is that accounting for large, yet realistic, levels of dispersion preserves the essential threshold-behaviour predicted by completely PT-symmetric dispersionless models, while more moderate levels of dispersion can completely change the character of the response of the system to increasing gain and loss. In particular, when there is moderate dispersion the gain and loss materials respond differently to frequency shifts in such a way that sharp threshold points give way to gradual changes over a range of parameters. When dispersion is increased further, the response reverts to threshold behaviour of the type seen in non-dispersive PT-symmetric systems, albeit with some breaking of detailed quantitative symmetry.

Our recent work on dispersive PT-Bragg gratings has shown that material dispersion limits the unidirectional invisibility to a single frequency which is in stark contrast to previous results that assumed idealized gain/loss profile in order to demonstrate wideband unidirectional behavior\cite{phang14a,phang14b}. In this paper, the performance of the microresonator-based PT system is analyzed for practical scenarios involving: a) frequency mismatch between the cavity resonant frequency and the gain pump frequency and, b) imperfect balance of the gain and loss in the system. The analysis of the microresonator-based PT system is achieved using an exact representation of the problem based on boundary integral equations and explicit analytical results are given for a weakly coupled system using perturbation analysis\cite{creagh01}. We concentrate on the weakly-coupled limit in our detailed calculations because that captures the essential properties of the threshold behavior of the PT-symmetry while allowing simple analytical calculations to be used. Finally, real-time field evolution in a two microresonator PT-symmetric system is analyzed for different levels of dispersion using the numerical time domain Transmission Line Modelling (TLM) method\cite{christopulos,paul99,phang13}. 

\section{PT symmetric coupled microresonators}

\begin{figure}[tbp]
\centering
\includegraphics[width=0.5\linewidth]{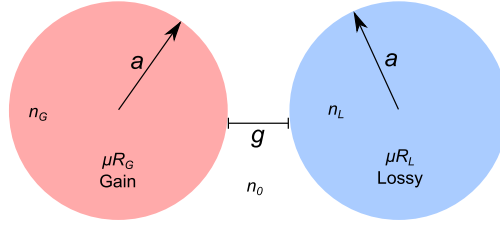}
\caption[Schematic of two coupled cylindrical microresonators of radius $a$ and separated by a distance $g$. $\mu R_G$ and $\mu L_G$ denote gain and lossy microresonator, respectively.]{Schematic of two coupled cylindrical microresonators or radius a and separated by a distance g. Microresonators with gain and loss are denoted by $\mu R_G$ and $\mu R_L$, respectively.}
\label{fig:illus}
\end{figure}

In this section we describe the theoretical background of a PT-symmetric system based on two coupled microresonators. The system, in which both microresonators have radius $a$ and are separated by a gap $g$, is illustrated schematically in Fig. \ref{fig:illus}. The active and passive microresonators have complex refractive indices $n_G$ and $n_L$ respectively, that are typically chosen to satisfy the PT condition $n_G = n_L^*$, where ‘*’ denotes complex conjugate, $n=(n'+jn'')$, and $n'$ and $n''$ represent the real and imaginary parts of the refractive index. In practice, localized gain might be achieved by means of erbium doping and optical pumping of the active microresonator, while masking the lossy microresonator as in\cite{ruter10,peng14,feng14,peng14b}. Both resonators are assumed to be surrounded by air. 

The refractive index of dispersive materials is frequency dependent but must also satisfy the causality property between  the real and imaginary parts of the material refractive index\cite{landau,zyablovsky14}. The material properties are conveniently modelled by assuming a dielectric constant that uses a Lorentzian model for dispersion as in\cite{hagness}
\begin{equation}
\varepsilon_r(\omega) = \varepsilon_\infty-j\frac{\sigma_0}{2\varepsilon_0\omega}\left(\frac{1}{1+j(\omega+\omega_\sigma)\tau} + \frac{1}{1+j(\omega-\omega_\sigma)\tau} \right). 
\end{equation}
Here $\varepsilon_\infty$ denotes the permittivity at infinity, $\omega_\sigma$ denotes the atomic transitional angular frequency, $\tau$ is the dipole relaxation time and $\sigma_0$ is related to the conductivity peak value that is set by the pumping level at $\omega_\sigma$. The time-varying component has been assumed to be of the form $e^{j\omega t}$ and therefore $\sigma_0>0$ denotes loss while $\sigma_0<0$ denotes gain. The parameter $\tau$ controls the degree of dispersion, with $\tau=0$ corresponding to a dispersion-less system. Throughout this paper, the frequency-dependent refractive index is expressed as $n=\sqrt{\varepsilon_r(\omega)}$ and the material gain/loss parameter is expressed using the imaginary part of refractive index as $\gamma=\omega n^{''}$. 

\section{Analysis of inter-resonator coupling in the frequency domain}
We now provide an analysis of coupling between resonators based on boundary integral methods. This approach is particularly suited to perturbative approximation of the coupling strength in the weak coupling limit but also provides an efficient platform for exact calculation when coupling is strong. The calculation is based on an approach used in\cite{creagh01} to describe coupling between fully bound states in coupled resonators and optical fibers, but adapted here to allow for radiation losses. It is also similar to methods used in\cite{smotrova06,smotrova13,boriskina06}.

\subsection{Notation and assumptions}
We assume resonators of radius $a$, uniform refractive index and TM boundary conditions. Then the mode taking the form $\psi_L=(J_m(n_Lkr)/J_m(n_Lka))e^{jm\theta}$ inside the isolated lossy  resonator is such that the solution and its normal derivative on the boundary of the resonator can be written as
\begin{equation}\label{useF}
a\dydxv{\psi_L}{n} =  F^{L}_m \psi_L,
\end{equation}
where 
\begin{equation}\label{defF}
F^{L}_m = \frac{zJ_m'(z)}{J_m(z)} \qquad\mbox{and \qquad $z=n_{L}ka$},
\end{equation}
where, $k$ is the free-space wave number and $\psi_G$ and $F^G_m$  being defined similarly for the gain resonator. The treatment of coupling in the remainder of this section can be used for other circularly-symmetric resonators such as those with graded refractive index or with different boundary conditions, as long as an appropriately modified $F^L_m$ is substituted in (\ref{useF}). 

\subsection{Exact solution using boundary-integral representation }
An exact boundary integral representation of the coupled problem is conveniently achieved by applying Green's identities to a region $\Omega$ which excludes the resonators, along with an infinitesimally small layer surrounding them (so that the boundaries $B_G$ and $B_L$ of the resonators themselves lie just outside $\Omega$). In $\Omega$, we assume that the refractive index takes the value $n_0=1$, so that the free-space Green's function is
\begin{equation}
G_0(\x,\x') = -\frac{j}{4} H_0 (k|\x-\x'|),
\end{equation}
where $H_0(z) = J_0(z)-jY_0(z)$ denotes the Hankel function of the second kind (and the solution is assumed to have time dependence $\e^{j\omega t}$). Then, applying Green's identities to the region $\Omega$ and assuming radiating boundary conditions at infinity leads to the equation
\begin{equation}\label{Green}
0 = \int_{B_G+B_L}
\left(G_0(\x,\x')\dydxv{\psi(\x')}{n'}-
\dydxv{G_0(\x,\x')}{n'}\psi(\x')\right)\d s'
\end{equation}
when $\x$ lies on either $B_L$ or $B_G$ (and therefore just outside of $\Omega$).

We now expand the solution on the respective resonator boundaries as Fourier series, 
\begin{equation}
\psi_G = \sum_m \alpha_m^G \e^{jm\theta_G}
\qquad\mbox{and}\qquad
\psi_L = \sum_m \alpha_m^L \e^{jm\theta_L},
\end{equation}
in the polar angles $\theta_G$ and $\theta_L$ centered respectively on the gain and lossy resonators, running in opposite senses in each resonator and zeroed on the line joining the two centers. The corresponding normal derivatives can be written 
\begin{equation}
\dydxv{\psi_G}{n} = \sum_m \frac{1}{a} F_m^G \alpha_m^G \e^{jm\theta_G}
\qquad\mbox{and}\qquad
\dydxv{\psi_G}{n} = \sum_m \frac{1}{a} F_m^G \alpha_m^G \e^{jm\theta_G}.
\end{equation}
Using Graf's theorem\cite{abramowitz} to expand the Green's function $G_0(\x,\x')$ analogously in polar coordinates on each boundary, the integral equation (\ref{Green}), evaluated separately for $\x$ on $B_L$ and on $B_G$, leads to a pair of matrix equations
\begin{eqnarray}\label{sys1} 
D^G\alpha^G+ C^{GL} \alpha^L&=& 0\nonumber\\ 
C^{LG}\alpha^G+ D^L \alpha^L&=& 0.
\end{eqnarray} 
Here, 
\begin{equation}
\alpha^G = \left(\begin{array}{c}\vdots\\\alpha_m^G\\\alpha_{m+1}^G\\\vdots\end{array}\right)
\qquad\mbox{and}\qquad
\alpha^L = \left(\begin{array}{c}\vdots\\\alpha_m^L\\\alpha_{m+1}^L\\\vdots\end{array}\right)
\end{equation}   
are Fourier representations of the solution on the boundaries of the gain and lossy resonators respectively. The matrices $D^G$ and $D^L$ are diagonal with entries 
\begin{equation}
D^{G,L}_{mm} =
J_m(u)H_m(u)\left(F_m^{G,L}-\frac{uH_m'(u)}{H_m(u)}\right),
\qquad\mbox{where $u = ka$},
\end{equation}
and provide the solutions of the isolated resonators.  The matrices
$C^{GL}$ and $C^{LG}$ describe coupling between the resonators. The matrix
$C^{GL}$ has entries of the form
\begin{equation}
C_{lm}^{GL} = J_l(u)H_{l+m}(w)
J_m(u)
\left(F_m^{L}-\frac{u_L J_m'(u)}{J_m(u)}\right),
\end{equation}
where $u=ka$, $w = kb$ and $b$  is the center-center distance between the gain and lossy resonators. The matrix $C^{LG}$ is defined analogously by exchanging the labels $G$ and $L$.

\subsection{PT-symmetry in the exact solution}
The system (\ref{sys1}) can be presented more symmetrically by using the
scaled Fourier coefficients
\begin{equation}
\tilde{\alpha}^L_m =
J_m(u)
\left(F_m^L-\frac{uJ_m'(u)}{J_m(u)}\right)
\alpha_m^L
\end{equation}
(along with an analogous definition of $\tilde{\alpha}_m^G$). Then 
(\ref{sys1}) can be rewritten
\begin{eqnarray}\label{sys2} 
\tilde{D}^G\tilde{\alpha}^G+ \tilde{C} \tilde{\alpha}^L&=& 0\nonumber\\ 
\tilde{C}\tilde{\alpha}^G+ \tilde{D}^L \tilde{\alpha}^L&=& 0,
\end{eqnarray} 
where the diagonal matrices $\tilde{D}^{G,L}$ have entries
\begin{equation}
\tilde{D}^{G,L}_{mm} =
-j \frac{H_m(u)F_m^{G,L}-uH_m'(u)}{J_m(u)F_m^{G,L}-uJ_m'(u)},
\qquad\mbox{where $u = ka$},
\end{equation}
and the same (symmetric) matrix $\tilde{C}$, with entries
\begin{equation}
\tilde{C}_{lm} =-j H_{l+m}(w),
\end{equation}
couples solutions in both directions. 

We have included an overall factor of $-j$ in these equations to emphasise an 
approximate PT-symmetry that occurs when
$n_G=n_L^*$. Then, in the limit of high-$Q$ whispering gallery
resonances for which we may approximate
\begin{equation}
jH_m(u)\simeq Y_m(u)\qquad\mbox{and}\qquad
jH_{l+m}(u)\simeq Y_{l+m}(u),
\end{equation}
the matrices in (\ref{sys2}) satisfy the conditions
\begin{equation}
\left(\tilde{D}^L\right)^* \simeq \tilde{D}^G 
\qquad\mbox{and}\qquad
\tilde{C}^* \simeq \tilde{C}
\end{equation}
which are a manifestation of PT symmetry of the system as a
whole: deviation from  these conditions is due to radiation
losses.

\subsection{Perturbative weak-coupling approximation}
The system of equations (\ref{sys2}) can be used as the basis of an
efficient numerical method for determining the resonances of the
coupled system with arbitrary accuracy. In practice, 
once the gap $g=b-2a $ between the resonators is  
wavelength-sized or larger,
a truncation of the full system to a relatively 
small number of modes suffices to describe the full solution.

In the limit of very weak coupling we may achieve an effective 
perturbative approximation by restricting our calculation to 
a single mode in each resonator. We consider in particular the case of
near left-right symmetry in which 
\begin{equation}
n_G\approx n_L. 
\end{equation}
PT symmetry is achieved by further imposing $n_G=n_L^*$, but for
now we allow for the effects of dispersion by not assuming that this is
the case. We build the full solution around modes for which
\begin{equation}
\psi_{\pm} \approx \psi_G \pm \psi_L,
\end{equation}
where $\psi_G$ and $\psi_L$ are the solutions of the isolated
resonators described at the beginning of this section. We 
use a single value of $m$ for both $\psi_G$ and $\psi_L$ and in 
particular approximate the global mode using a chiral state in 
which the wave circulates in opposite senses in each resonator.
That is, we neglect the coupling between $m $ and $-m$ that occurs in
the exact solution.

Then a simple perturbative approximation is achieved by
truncating the full system of equations (\ref{sys2}) to the $2\times 2$ system
\begin{equation}
M
\left(
\begin{array}{c}\tilde{\alpha}^G_{mm}\\ \tilde{\alpha}^L_{mm}\end{array}
\right) = 0,
\qquad\mbox{where}\quad
M=\left(
\begin{array}{cc}\tilde{D}^G_{mm}&\tilde{C}_{mm}\\\tilde{C}_{mm}&\tilde{D}^L_{mm}\end{array}
\right).
\end{equation}
Resonant frequencies of the coupled problem are then realised when
\begin{equation}
0 = \det M = \tilde{D}^G_{mm}\tilde{D}^L_{mm}-\tilde{C}_{mm}^2.
\end{equation}
In the general, dispersive and non-PT-symmetric, case this reduces the
calculation to a semi-analytic solution in which we search for the
(complex) roots of the known $2\times 2$ determinant above, in which matrix elements depend on
frequency through both $k=\omega/c$ and $n=n(\omega)$.

\subsection{Further analytic development of the perturbative solution}
To develop a perturbative expansion we let 
\begin{equation}
D_{mm}^0 = \frac{1}{2}\left(\tilde{D}_{mm}^G+\tilde{D}_{mm}^L\right) 
\qquad \mbox{and} \qquad
D_{mm}^I = \frac{1}{2j}\left(\tilde{D}_{mm}^G-\tilde{D}_{mm}^L\right)
\end{equation} 
(and note that in the high-$Q$-factor PT-symmetric case,
$\tilde{D}^G\simeq(\tilde{D}^L)^*$, both $D_{mm}^0$ and $D_{mm}^I$ are
approximately real). We assume that both $D^I_{mm}$ and $C_{mm}$ are
small and comparable in magnitude, and expand angular frequency
\begin{equation}
\omega_{1,2} = \omega_0 \pm \frac{\Delta\omega_0}{2} + \cdots
\end{equation}
about a real resonant angular frequency of an averaged isolated
resonator satisfying
\begin{equation}
D^0_{mm}(\omega_0) = 0.
\end{equation}
Then to first order the coupled resonance condition becomes
\begin{equation}
0=\det M = \Delta\omega_0^2 {D_{mm}^0}'(\omega_0)^2 +
D_{mm}^I(\omega_0)^2-\tilde{C}_{mm}(\omega_0)^2 + \cdots
\end{equation}
from which the angular frequency shifts can be written
\begin{equation}\label{getomega1}
\frac{\Delta\omega_0}{2} = \frac{\sqrt{\tilde{C}_{mm}(\omega_0)^2-D_{mm}^I(\omega_0)^2}}{{D_{mm}^0}'(\omega_0)},
\end{equation}
where ${D_{mm}^0}'(\omega)$ denotes a derivative of $D_{mm}^0(\omega)$
with respect to frequency.

We then arrive at a simple condition
\[
\tilde{C}_{mm}(\omega_0)^2=D_{mm}^I(\omega_0)^2 
\]
for threshold (at which $\Delta\omega_0$ and the two resonant frequencies
of the coupled system collide). In the PT-symmetric case, where
$\tilde{C}_{mm}$ and $D_{mm}^I$ are approximately real (and whose
small 
imaginary parts represent corrections due to radiation 
losses), we therefore have a prediction for a real threshold frequency.

\section{Results and discussions}
In this section, the impact of dispersion on the performance of the PT coupled microresonators is analyzed. Frequency mismatch between the resonant frequency of the microresonator and gain pump frequency is investigated for practical levels of dispersion and the practical implications of a slight unbalance between the gain and loss in the system are investigated.  We conclude the section with an investigation of how coupling between resonators manifests itself in the time development of solutions.

\subsection{Effects of dispersion on threshold behavior in the frequency domain.} 
For all cases, weakly coupled microresonators are considered, the coupled resonators with a dielectric constant $\varepsilon_\infty
=3.5$\cite{hagness}  have radius $a = 0.54 \mu\mbox{m}$ and are separated by distance $g = 0.24 \mu\mbox{m}$. Transverse-magnetic (TM) polarization is considered and operation at two different whispering-gallery modes is analysed, namely a low $Q$-factor mode (7,2) and a high $Q$-factor mode (10,1). The corresponding isolated resonator resonant frequencies are respectively $f_0^{(7,2)}=341.59\mbox{THz}$ and $f_0^{(10,1)}=336.85\mbox{THz}$, with $Q$-factors $Q^{(7,2)}=2.73\times10^3$ and $Q^{(10,1)}=1.05\times10^7$. 

\begin{figure}
\centering
\includegraphics[width=0.9\linewidth]{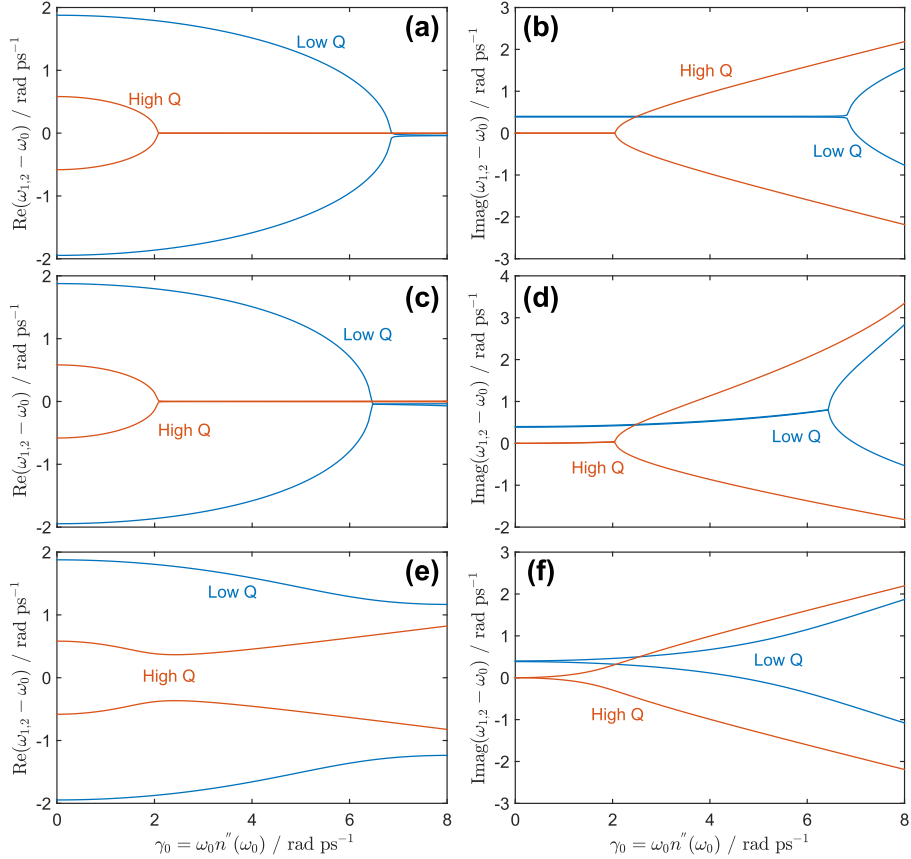}
\caption{Frequency bifurcation of PT-coupled microresonator with a balanced gain ($\gamma_G=-\gamma_0$) and loss ($\gamma_L=\gamma_0$) as a function of gain/loss parameter at the peak of pumping beam $\gamma_0=\omega_{\sigma}n^{''}(\omega_{\sigma}$) for three different dispersion parameters, (a,b) $\omega\tau=0$, (c,d) $\omega\tau=212$ and (e,f) $\omega\tau=0.7$}
\label{fig:spliting}
\end{figure}

Figure \ref{fig:spliting} shows the real and imaginary part of the eigenfrequencies $\omega_1$ and $\omega_2$ of the PT-symmetric coupled microresonators with balanced gain and loss, $\gamma_0=-\gamma_G=\gamma_L$, and is depicted as a function of the gain/loss parameter $\gamma_0=|\omega_0n^{''}(\omega_0)|$ for both the low and high $Q$-factor modes. The gain and loss are assumed to be tuned to the resonant frequency of an isolated microresonator, i.e. $\omega_\sigma=\omega_0\equiv2\pi f_0$. Three different levels of dispersion, controlled by the parameter $\tau$ defined in Sec. 2, are considered. These are $\omega_\sigma\tau=0$ corresponding to the case of no dispersion, $\omega_\sigma\tau=212$ taken from\cite{hagness} to exemplify the case of high dispersion and $\omega_\sigma\tau=0.7$ to exemplify the case of low dispersion.

Figure \ref{fig:spliting}(a,b) shows the frequency splitting of the real and imaginary part of the complex eigenfrequencies for the case of no dispersion. In the absence of gain/loss, where $\gamma_0=0$, the supermodes beat at a rate corresponding to the frequency differences $\omega_1-\omega_2=3.823\mbox{ rad/ps}$ and 1.164 rad/ps for the (7,2) and (10,1) modes respectively. Figure \ref{fig:spliting}(a) indicates that operation in a higher $Q$-factor mode results in weaker coupling between the microresonators compared to the case of operation in the lower $Q$-factor mode. Increasing the gain and loss in the system, decreases the beating rate and the supermodes coalesce at the threshold points of $\gamma_0=6.86\mbox{ rad/ps}$ and 2.1 rad/ps for the low and high $Q$-factor modes of operation respectively, confirming that the high-$Q$ factor mode has a lower threshold point\cite{peng14}. In the case of operation in the low $Q$-factor mode, the eigenfrequencies shown in Fig. \ref{fig:spliting}(b) have a significant constant and positive imaginary part before the threshold point, which is a consequence of the higher intrinsic losses due to radiation in that case. The corresponding imaginary part is insignificant in the case of the high Q-factor mode, for which radiation losses are much smaller. Furthermore it is noted here that the coupled system first starts to lase, i.e. one of the eigenfrequencies satisfies $\mbox{Im}(\omega_{1,2}-\omega_0)<0$, only when operated significantly beyond the threshold $\gamma_0=7\mbox{ rad/ps}$ for the low $Q$-factor operation while this onset occurs immediately after the threshold point in the high $Q$-factor case. 

Figure \ref{fig:spliting}(c,d) shows the real and imaginary parts of the eigenfrequencies for the case of strong dispersion, corresponding to the parameter values $\omega_\sigma\tau=212$ taken from\cite{hagness}. These are again shown for both high and low $Q$-factor modes. It is noted that the threshold point for the low $Q$-factor mode is reduced from  $\gamma_0=6.86\mbox{ rad/ps}$ to  $\gamma_0=6.47\mbox{ rad/ps}$ in this case while for the high $Q$-factor mode it remains unchanged at 2.1 rad/ps (compared to the case of no dispersion). Below the threshold point the imaginary parts of the eigenfrequencies are not constant, but are instead skewed towards a lossy state with positive and increasing imaginary part. Extension beyond the threshold point shows that the imaginary parts of the eigenfrequencies do not split evenly and are also skewed towards overall loss, implying that in the highly dispersive case the eigenfrequencies both are complex but no longer complex conjugates after the threshold point. 

The real and imaginary parts of the eigenfrequencies for the case of low levels of dispersion, for which we take $\omega_\sigma\tau=0.7$, are shown in Fig. \ref{fig:spliting}(e,f). Figure \ref{fig:spliting}(e) shows that there is no clear threshold point in this case: the imaginary parts split for very low value of the gain/loss parameter $\gamma_0$, with no sharp point of onset. The appearance of a threshold point typically associated with PT-behavior is lost and the eigenfrequencies are always complex valued. 

The key conclusion to be made from Fig. \ref{fig:spliting} is therefore that PT-like threshold behavior is observed in the cases of no dispersion and of high dispersion, but not for cases of intermediate dispersion. While there is some skewness in the high-dispersion case, which amounts to a quantitative deviation from strict PT-symmetry, there is an essential qualitative similarity to the dispersionless case in which there appears to be a sharp threshold. By contrast, in the case of intermediate dispersion there is no sharp transition point and the imaginary parts of the two frequencies begin to diverge from the beginning. 

\begin{figure}
\centering
\includegraphics[width=0.9\linewidth]{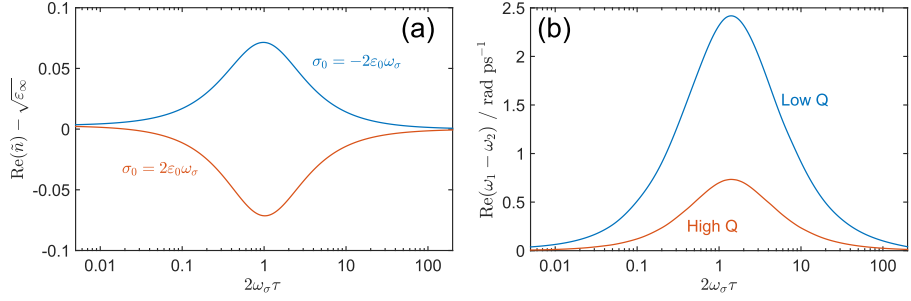}
\caption{(a) Impact of dispersion to the real part of material at atomic transitional angular frequency $\omega_\sigma$ due to the presence of gain and loss for different dispersion parameters; (b) Contrast between the real part of eigenfrequencies of PT-coupled microresonators for two different gain/loss parameter, i.e. $\gamma_0=7.5\mbox{ rad/ps}$ for (7,2) and 2.54 rad/ps for the (10,1) mode as function of dispersion parameter $\tau$.}
\label{fig:mouth}
\end{figure}

To further investigate and explain this phenomenon, we examine the dependence of the real part of the complex refractive index on the dispersion parameter $2\omega_\sigma\tau$. This dependence is plotted in Fig. \ref{fig:mouth}(a) for the cases of both gain and loss, for which we respectively take $\sigma_0=\pm2\varepsilon_0\omega_\sigma$ and $\omega_\sigma=\omega_0$. Figure \ref{fig:mouth}(a) shows that the real parts of the refractive indices behave differently for the cases of loss and gain in the system, with the maximum difference occurring when $\tau=1/(2\omega_\sigma)$. However, in two limiting cases $\tau=0$ (dispersion-less system) and $\tau\rightarrow\infty$ (strong dispersion), the real parts of the refractive index converge.  This means that the PT condition $n_G=n_L^*$  can only be satisfied accurately for the cases of no dispersion and of high dispersion. For the case of intermediate dispersion there is necessarily some discrepancy between the real parts of the refractive indices of the resonators.

\begin{figure}
\centering
\includegraphics[width=0.9\linewidth]{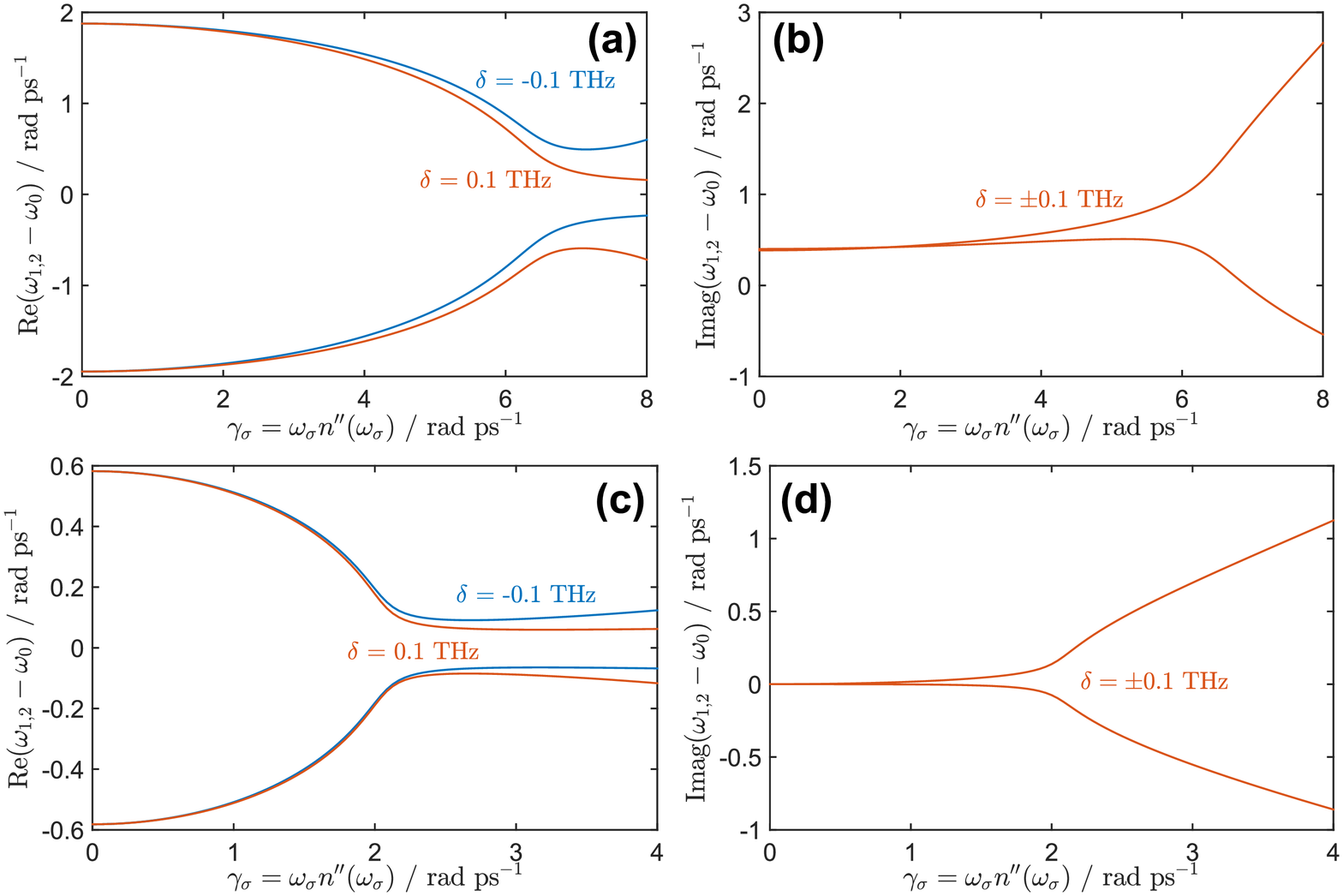}
\caption{Frequency bifurcation of coupled microresonators with balanced gain and loss as function of gain/loss parameters $\gamma_\sigma$, for two different atomic transitional frequencies $\omega_\sigma=2\pi(f_0+\delta)$ with $\delta =-0.1$ and 0.1 THz.}
\label{fig:mismatch}
\end{figure}

Figure \ref{fig:mouth}(b) shows the minimum difference in the real parts of the two eigenfrequencies for different dispersion levels and operated at a fixed value of the gain/loss parameter, i.e. at  $\gamma_0=7.5\mbox{ rad/ps}$ for the low $Q$-factor and at 2.54 rad/ps for high $Q$-factor modes of operation. These values of the gain/loss parameter are chosen to lie above the expected threshold so that qualitatively PT-like behavior would imply eigenfrequencies with a common real part. Figures \ref{fig:mouth}(a) and \ref{fig:mouth}(b) confirm that the maximum difference between real parts of the two refractive indices coincides with the maximum deviation from PT-like threshold behavior, where the difference between real parts of the eigenfrequencies is greatest. This result further confirms the fact that realistic levels of dispersion preserve the essential features of PT behavior. 

Having confirmed that realistic levels of dispersion preserve PT behavior, Figure \ref{fig:mismatch} considers a practical scenario in which there is high dispersion $\omega_\sigma\tau=212$ and a frequency mismatch between the resonant frequency and the gain/loss atomic angular frequency. The material atomic frequency is defined to be $\omega_\sigma=2\pi(f_0+\delta)$, where $\delta$ is the mismatch parameter. The structure is operated with balanced gain and loss, i.e. $\gamma_G=-\gamma_L$ and two values are assumed for the frequency mismatch, namely $\delta=-0.1$ and 0.1 THz. Figure \ref{fig:mismatch}(a,b) shows the results for the low $Q$-factor mode (7,2) and Fig. \ref{fig:mismatch}(c,d) for the high-$Q$ factor mode (10,1). In both cases there is no sharp threshold point for the real parts of eigenfrequencies and the imaginary parts begin to diverge at low gain/loss values. Neither are the imaginary parts symmetrically placed about a branching value. This result confirms the fact that PT behavior is preserved only when the angular transitional frequency of the dispersive gain/loss profile is aligned with the resonant frequency of the microresonators. If that is not the case, the frequency misalignment causes the coupled system to continue to beat after a threshold region.

\begin{figure}[h]
\centering
\includegraphics[width=0.9\linewidth]{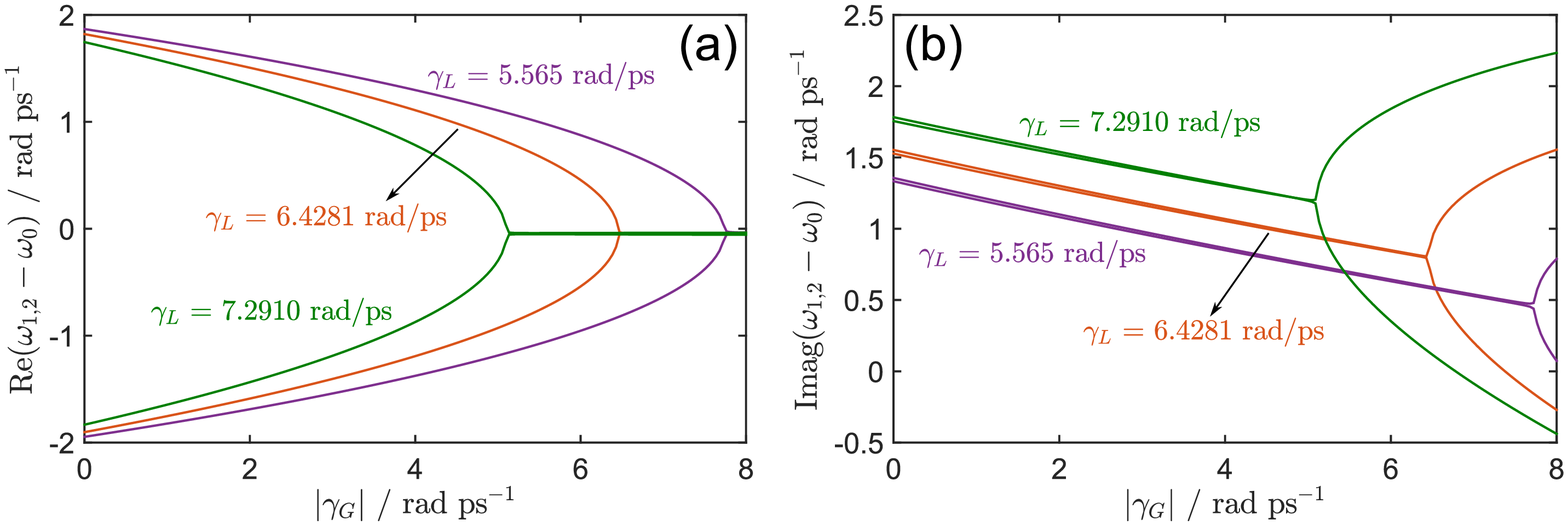}
\caption{Complex eigenfrequency in a PT-coupled microresonator system with variable gain and fixed loss shown as a function of gain parameter $|\gamma_G|$, dispersion parameter $\omega_\sigma\tau=212$\cite{hagness} and shown for 3 different fixed loss value, i.e. $\gamma_L=5.565$, 6.4281, and 7.291 rad/ps.}
\label{fig:vargain}
\end{figure}

Another practical scenario is considered in Fig. \ref{fig:vargain} where the gain and loss are not balanced, i.e.  $\mu R_L$ has a loss $\gamma_L$ while $\mu R_G$ has a gain $\gamma_G$. Figure \ref{fig:vargain}(a,b) shows the real and imaginary parts of the eigenfrequency for three different values of loss namely, $\gamma_L=5.565$, 6.4281 and 7.291 rad/ps which correspond to values below, at, and above the threshold point of a PT symmetric structure with balanced gain and loss respectively. The low $Q$-factor mode is considered with a practical dispersion parameter of $\omega_\sigma\tau=212$ as taken from \cite{hagness}. Interestingly, we now observe that the PT threshold point can also exist for structures with unbalanced gain/loss as shown by the plots for $\gamma_L=5.565\mbox{ rad/ps}$ and $\gamma_L=7.2910\mbox{ rad/ps}$ in Fig. \ref{fig:vargain}. In the former case, the PT threshold is increased and in the latter case the PT threshold is decreased compared to the PT threshold of the balance structure. Of special interest is the observation that increasing loss results in the reduction of the PT threshold which consequently reduces the levels of gain at which lasing occurs. This counter-intuitive principle of switching lasing on by increasing loss has been experimentally demonstrated in \cite{peng14b} where a metal probe is used to enhance loss in the lossy microresonator. 

\subsection{Real time operation of PT symmetric coupled microresonators}
In this section the real-time operation of the PT-symmetric coupled microresonators is demonstrated for different levels of dispersion. For this purpose, the two-dimensional (2D) time-domain Transmission Line Modeling (TLM) numerical method is used.  A more detailed description of the TLM method is given in\cite{christopulos,paul99} and the implementation of general dispersive materials for PT-symmetric Bragg gratings is demonstrated in\cite{phang14a,phang15}. In each of the simulations shown in this section, the low $Q$-factor (7,2) mode is excited by a very narrow-band Gaussian dipole located in $\mu R_G$ whose frequency is matched to the resonant frequency of this mode. Depending on the levels of gain and loss, and their relation to the threshold points, we find in practice, however, that small unintentional initial excitations of the high $Q$-factor (10,1) mode may grow to become a significant feature and even dominate the evolved state. In all cases we find that the TLM simulations are consistent with the frequency-domain calculations provided in the previous section and in fact have been used to independently validate the perturbation analysis results presented in Figs. \ref{fig:spliting}-\ref{fig:vargain}, although the detailed calculations are not reported here.

\begin{figure}
\centering
\includegraphics[width=0.9\linewidth]{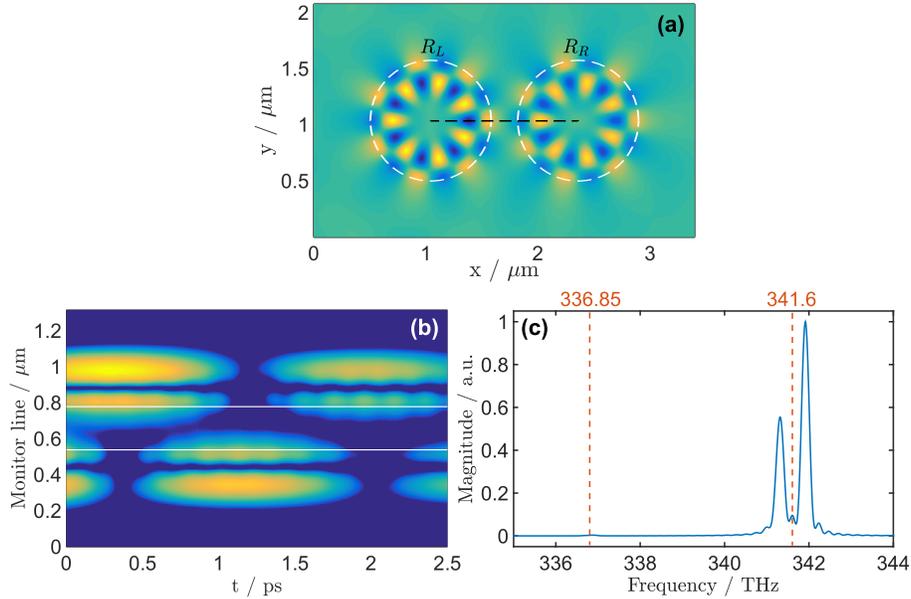}
\caption{(a) Spatial electric field distribution of the coupled microresonators operated (7,2) mode. The black dashed line denotes the monitor line. The temporal evolution (b) and spectra (c) of the field on the monitor line are shown in the absence of gain and loss.}
\label{fig:tlmnogain}
\end{figure}
\begin{figure}
\centering
\includegraphics[width=0.9\linewidth]{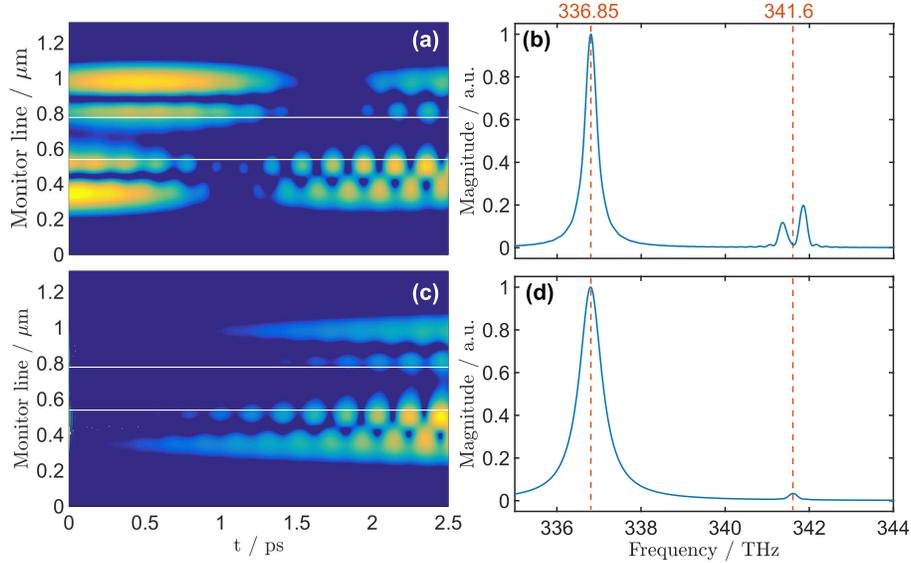}
\caption{The temporal evolution and spectra of the field on the monitor line are shown for different gain/loss parameters, (a,b) for $\gamma_0 = 4.3\mbox{ rad/ps}$ and (c,d) for $\gamma_0 = 7.5\mbox{ rad/ps}$ with a negligible dispersion parameter using the TLM method.}
\label{fig:nodispersion}
\end{figure}

We begin with the case of the evolution from the low $Q$-factor mode using a model with no dispersion. Figure \ref{fig:tlmnogain}(a) shows the spatial electric field distribution of coupled microresonators with no gain and loss ($\gamma_0 = 0$) and operating at the resonant frequency of the low $Q$-factor (7,2) mode. The black dashed line denotes the monitor line on which the electric field is observed during the TLM simulation. Parts (b,c) of Fig. \ref{fig:tlmnogain} show the temporal evolution and the spectra of the electric field observed along the monitor line for the case of no gain and loss. 

The case of no gain and loss, reported in Fig. 6(b), shows a typical oscillation of the electric field between the microresonators having a regular beating pattern in which maximum intensity being observed in one microresonator corresponds to minimum intensity being observed in the other. Figure 6(c) shows the frequency content of the modes, indicating the presence of two resonating frequencies centered around $f_0^{(7,2)}$, in agreement with Fig. 2(a). 

The real-time performance of PT-coupled microresonators with balanced gain and loss $\gamma_G=-\gamma_0$, $\gamma_L=\gamma_0$ with no dispersion is depicted in Fig. \ref{fig:nodispersion}. The temporal evolution and the spectra of the electric field are observed along the monitor line for two levels of gain and loss: levels of gain and loss $\gamma_0 = 4.3\mbox{ rad/ps}$, i.e. lower than the threshold point of the low $Q$-factor mode but beyond the threshold of the high Q-factor mode in Fig. \ref{fig:nodispersion} (a and b) and levels of gain and loss $\gamma_0 = 7.5\mbox{ rad/ps}$, i.e. lying above the threshold points of both modes in Fig. \ref{fig:nodispersion} (c and d).

Figure \ref{fig:nodispersion}(a) shows the energy beating between the microresonators with gain/loss $\gamma_0 = 4.3\mbox{ rad/ps}$, set below the threshold point of the low-$Q$-factor mode. It is noticeable that beating between microresonators is reduced and is no longer periodic. Additionally, Fig. \ref{fig:nodispersion}(a) indicates the presence of additional modes, observable at later times. Frequency analysis of the fields is given in Fig. \ref{fig:nodispersion}(b) and shows an additional peak at 336.85 THz, which corresponds to the resonant frequency of the mode (10,1), explaining the high frequency beating in Fig. \ref{fig:nodispersion}(a). Referring to Fig. \ref{fig:spliting}(a) it can be seen that at $\gamma_0 = 4.3\mbox{ rad/ps}$, the (10,1) mode is operating above its threshold point and thus experiencing amplification, whilst the (7,2) mode is still below its threshold point. 

A further increase in gain/loss to $\gamma_0 = 7.5\mbox{ rad/ps}$, in Fig. \ref{fig:nodispersion}(c), shows an exponentially growing field in the gain microresonator with no beating between the resonators and a stronger presence of the high $Q$-factor mode as shown in Fig. \ref{fig:nodispersion}(d). Referring to Fig. \ref{fig:spliting}(a,b) it is confirmed that for $\gamma_0 = 7.5\mbox{ rad/ps}$ both low and high $Q$-factor modes are operating above the threshold.  

\begin{figure}
	\centering
	\includegraphics[width=0.9\linewidth]{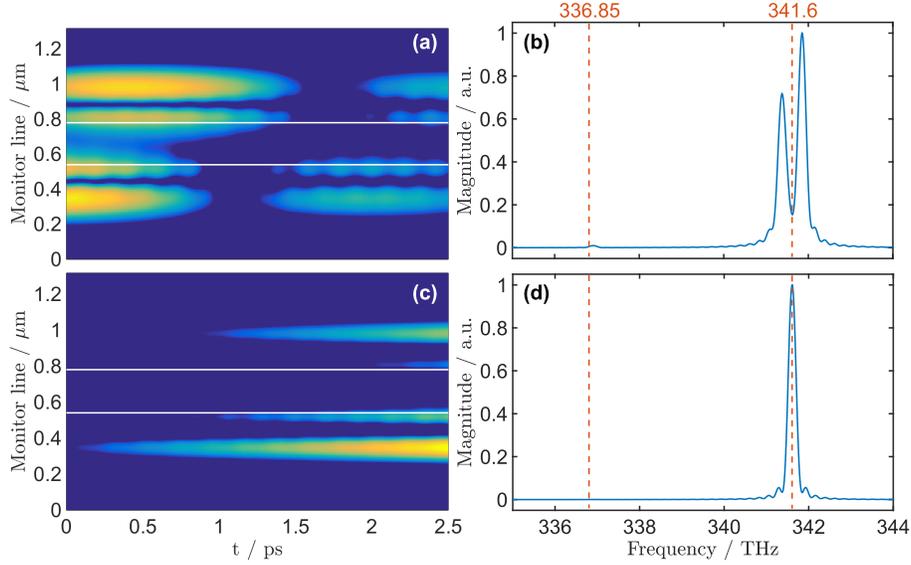}
	\caption{Temporal and spectra of electric field along the monitor line for coupled PT microresonators with balanced gain and loss parameters operated for (7,2) mode with practical dispersion parameters $\omega_\sigma\tau=212$\cite{hagness} and for two different gain/loss parameter, i.e. (a,b) $\gamma_0 = 4.3\mbox{ rad/ps}$ and (c,d) $\gamma_0 = 7.5\mbox{ rad/ps}$.}
	\label{fig:highdisp}
\end{figure}

Figure \ref{fig:highdisp} shows corresponding results for the case of high dispersion with  $\omega_\sigma\tau=212$. We once again choose gain/loss parameters  $\gamma_0 = 4.3\mbox{ rad/ps}$ and   $\gamma_0 = 7.5\mbox{ rad/ps}$ but omit the case of zero gain/loss here. Figure \ref{fig:highdisp}(a) shows a decaying beating pattern. The corresponding spectral analysis in fig. \ref{fig:highdisp}(b) shows that the beating may be attributed to low $Q$-factor modal frequencies, indicating that the highly dispersed gain/loss profile has stabilized the operation of PT-coupled resonators system at a desired mode of operation. For operation with gain/loss parameter  $\gamma_0 = 7.5\mbox{ rad/ps}$, the temporal response in Fig. \ref{fig:highdisp}(c) indicates an exponentially growing field with no presence of high order modes. The spectrum in Fig. \ref{fig:highdisp}(d) shows a single peak at the resonant frequency of the (7,2) mode, confirming that the resonators are operating above the PT threshold point.  The strongly dispersive gain/loss profile limits operation of PT-coupled microresonator system to the low $Q$-factor (7,2) mode only in this case.  This again confirms the result that when the material atomic frequency is chosen to be at a desired resonant frequency, the PT-symmetry is limited to that particular mode only, as in the case of periodic potentials \cite{phang14a,zyablovsky14}. 

\begin{figure}
\centering
\includegraphics[width=0.9\linewidth]{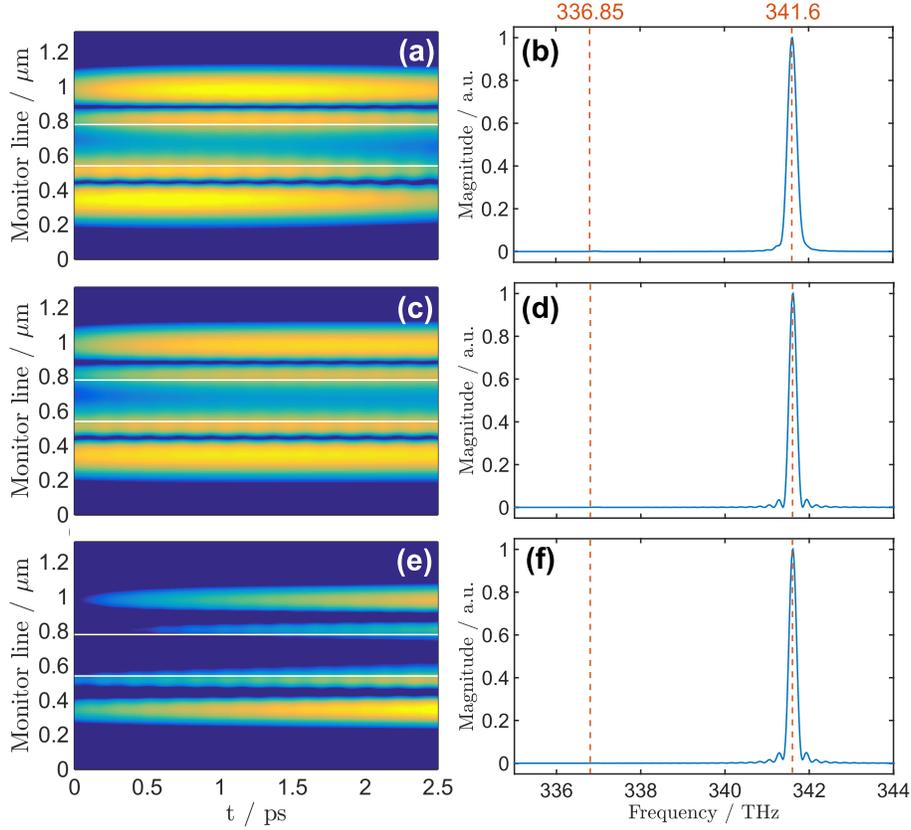}
\caption{Temporal and spectra of electric field along the monitor line for coupled PT microresonators with unbalanced gain and loss operated for (7,2) mode with practical dispersion parameters $\omega_\sigma\tau=212$\cite{hagness}, i.e. (a,b) $\gamma_L = 5.565\mbox{ rad/ps}$, (c,d) $\gamma_L = 6.4281\mbox{ rad/ps}$ and (e,f) $\gamma_L = 7.291\mbox{ rad/ps}$ while the gain parameter is kept constant at $\gamma_G = -7.053\mbox{ rad/ps}$.}
\label{fig:fixloss}
\end{figure}

Figure \ref{fig:fixloss} investigates the real-time operation of the PT structure with unbalanced gain and loss shown in Fig. \ref{fig:vargain}. Here, we apply the same scenario as in \cite{peng14b} where the gain in the active microresonator is fixed at $\gamma_G=-7.053\mbox{ rad/ps}$ and the loss is varied in the passive microresonator. The low Q-factor mode with practical dispersion parameters of $\omega_\sigma\tau=212$ is considerd \cite{hagness}. Figure \ref{fig:fixloss}(a) shows the electric field observed along the monitor line for the case of loss $\gamma_L=5.565\mbox{ rad/ps}$, i.e. more gain than loss in the system. It can be seen that there is a non-periodic and long temporal beating pattern. The field is not growing which indicates that the system is stable and is not lasing. The spectral decomposition shown in Fig. \ref{fig:fixloss}(b) is unable to distinguish the splitting of the resonant frequency is due to the limited spectral resolution of the Fourier transformation of the time domain simulation result. 

Figure \ref{fig:fixloss}(c) shows the temporal evolution of the field for the case of $\gamma_L=6.4281\mbox{ rad/ps}$. It can be seen that there is no beating between the resonators and no growing field, suggesting that structure is operating above the PT threshold point but before the lasing point which occurs at $\gamma_G=-7.377\mbox{ rad/ps}$ (Fig.\ref{fig:vargain}(b)). The spectral analysis shows  only a single peak centered at $f_0^{(7,2)}$.

Figure \ref{fig:fixloss}(e) shows the temporal evolution for $\gamma_L=7.291\mbox{ rad/ps}$ and $\gamma_G=-7.053\mbox{ rad/ps}$, i.e. more loss than gain in the system. It can be observed that the field is growing with no beating between the microresonators, suggesting lasing action. This result is in agreement with observations in \cite{peng14b} where loss induced lasing is demonstrated. Corresponding spectrum is depicted in Fig. \ref{fig:fixloss}(f) has a sharp peak centered at $f_0^{(7,2)}$.

\section{Summary and conclusion}
In the paper the impact of material dispersion on PT-symmetric coupled microresonators has been analyzed. It has been shown that the practical case of high dispersion preserves the requirement for a PT structure. However our results shows that this is only the case when the material atomic frequency is aligned with the resonant frequency of the microresonator. This comes as a direct consequence of the Kramers-Kronig relationship which implies that changes in the imaginary part of the refrective index caused the real part of the refractive index to change too. In addition, we also demonstrate the principle of loss-induced lasing mechanism which is triggered by an early PT-symmetry breaking. Real-time operation of PT-coupled microresonators verifies that the dispersion due to the Kramers-Kronig relationship limits the operation of PT-coupled microresonators to a single frequency and hence forbids multi-mode PT-symmetry breaking.  


\begin{thebibliography}{99}
\bibitem{bender99}C. M. Bender, S. Boettcher, and P. N. Meisinger, ``PT--symmetric quantum mechanics,'' J. Math. Phys. {\bf40}, 2201 (1999).

\bibitem{lin11} Z. Lin, H. Ramezani, T. Eichelkraut, T. Kottos, H. Cao, and D. N. Christodoulides, "Unidirectional invisibility induced by PT--iymmetric periodic structures," Phys. Rev. Lett. \textbf{106}, 213901 (2011).

\bibitem{chong11} Y. D. Chong, L. Ge, and A. D. Stone, ``PT--Symmetry breaking and laser-absorber modes in optical scattering systems," Phys. Rev. Lett. \textbf{106}, 093902 (2011).

\bibitem{benisty12} H. Benisty, C. Yan, A. Degiron, and A. Lupu, ``Healing near-PT-symmetric structures to restore their characteristic singularities: analysis and examples," J. Light. Technol. \textbf{30}, 2675--2683 (2012).

\bibitem{lupu13}A. Lupu, H. Benisty, and A. Degiron, ``Switching using PT symmetry in plasmonic systems: positive role of the losses," Opt. Express \textbf{21}, 192--195 (2013).

\bibitem{nazari11}F. Nazari, M. Nazari, and M. K. Moravvej-Farshi, ``A $2\times2$ spatial optical switch based on PT--symmetry.," Opt. Lett. \textbf{36}, 4368--4370 (2011).

\bibitem{hodaei14}H. Hodaei, M.-A. Miri, M. Heinrich, D. N. Christodoulides, and M. Khajavikhan, ``PT symmetric large area single mode DFB lasers," in \textit{CLEO: 2014} (OSA, 2014), Vol. 1, p. FM1D.3.

\bibitem{longhi14}S. Longhi and L. Feng, ``PT-symmetric microring laser-absorber.," Opt. Lett. \textbf{39}, 5026--5029 (2014).

\bibitem{phang13}S. Phang, A. Vukovic, H. Susanto, T. M. Benson, and P. Sewell, ``Ultrafast optical switching using parity--time symmetric Bragg gratings," J. Opt. Soc. Am. B \textbf{30}, 2984--2991 (2013).

\bibitem{phang15}S. Phang, A. Vukovic, T. M. Benson, H. Susanto, and P. Sewell, ``A versatile all--optical parity--time signal processing device using a Bragg grating induced using positive and negative Kerr-nonlinearity," Opt. Quantum Electron. \textbf{47}, 37--47 (2015).

\bibitem{kulishov13}M. Kulishov, B. Kress, and R. Slavík, ``Resonant cavities based on Parity-Time-symmetric diffractive gratings," Opt. Express \textbf{21}, 68--70 (2013).

\bibitem{jones12}H. F. Jones, ``Analytic results for a PT -symmetric optical structure," J. Phys. A Math. Theor. \textbf{45}, 135306 (2012).

\bibitem{ctyroky10}J. \v{C}tyrok\'{y}, V. Kuzmiak, and S. Eyderman, ``Waveguide structures with antisymmetric gain/loss profile," Opt. Express \textbf{18}, 21585--21593 (2010).

\bibitem{ctyroky14}J. \v{C}tyrok\'{y}, ``Dispersion properties of coupled waveguides with loss and gain: a full-vectorial analysis," Opt. Quantum Electron. \textbf{46}, 465--475 (2014).

\bibitem{ganainy07}R. El-Ganainy, K. G. Makris, D. N. Christodoulides, and Z. H. Musslimani, ``Theory of coupled optical PT-symmetric structures," Opt. Lett. \textbf{32}, 2632--2634 (2007).

\bibitem{greenberg04}M. Greenberg and M. Orenstein, ``Unidirectional complex grating assisted couplers.," Opt. Express \textbf{12}, 4013--4018 (2004).

\bibitem{longhi10}S. Longhi, ``PT-symmetric laser absorber," Phys. Rev. A \textbf{82}, 031801 (2010).

\bibitem{nolting96}H. Nolting, G. Sztefka, and J. \v{C}tyrok\'{y}, ``Wave propagation in a waveguide with a balance of gain and loss," in \textit{Integrated Photonics Research} (OSA, 1996), pp. 76--80.

\bibitem{phang14a}S. Phang, A. Vukovic, H. Susanto, T. M. Benson, and P. Sewell, ``Impact of dispersive and saturable gain/loss on bistability of nonlinear parity-time Bragg gratings.," Opt. Lett. \textbf{39}, 2603--2606 (2014).

\bibitem{ruter10}C. E. R\"{u}ter, K. G. Makris, R. El-Ganainy, D. N. Christodoulides, M. Segev, and D. Kip, ``Observation of parity–time symmetry in optics," Nat. Phys. \textbf{6}, 192--195 (2010).

\bibitem{peng14}B. Peng, \c{S}. K. \"{O}zdemir, F. Lei, F. Monifi, M. Gianfreda, G. L. Long, S. Fan, F. Nori, C. M. Bender, and L. Yang, ``Parity–time-symmetric whispering-gallery microcavities," Nat. Phys. \textbf{10}, 1--5 (2014).

\bibitem{peng14b} B. Peng, \c{S}. K. \"{O}zdemir, S. Rotter, H. Yilmaz, M. Liertzer, F. Monifi, C. M. Bender, F. Nori, and L. Yang, ``Loss-induced suppression and revival of lasing.," Science, \textbf{346}, no. 6207, pp. 328--332, (2014).

\bibitem{regensburger13}A. Regensburger, M. Miri, and C. Bersch, ``Observation of defect states in PT-symmetric optical lattices," Phys. Rev. Lett. \textbf{110}, 223902, (2013).

\bibitem{chang14}L. Chang, X. Jiang, S. Hua, C. Yang, J. Wen, L. Jiang, G. Li, G. Wang, and M. Xiao, ``Parity--time symmetry and variable optical isolation in active-passive-coupled microresonators," Nat. Photonics \textbf{8}, 524--529 (2014).

\bibitem{feng14}L. Feng, Z. J. Wong, R.-M. Ma, Y. Wang, and X. Zhang, ``Single-mode laser by parity-time symmetry breaking," Science \textbf{346}, 972--975 (2014).

\bibitem{phang14b}S. Phang, A. Vukovic, H. Susanto, T. M. Benson, and P. Sewell, ``Practical limitation on operation of nonlinear parity-time Bragg gratings," in \textit{META 2014 Conference} (2014), pp. 270--275.

\bibitem{creagh01}S. C. Creagh and M. D. Finn, ``Evanescent coupling between discs: a model for near-integrable tunnelling," J. Phys. A. Math. Gen. \textbf{34}, 3791--3801 (2001).

\bibitem{paul99}J. Paul, C. Christopoulos, and D. W. P. Thomas, ``Generalized material models in TLM .I. Materials with frequency-dependent properties," IEEE Trans. Antennas Propag. \textbf{47}, 1528--1534 (1999).

\bibitem{christopulos}C. Christopoulos, \textit{The Transmission-Line Modeling Method TLM} (IEEE Press, 1995).

\bibitem{landau}L. D. Landau, J. S. Bell, M. J. Kearsley, L. P. Pitaevskii, E. M. Lifshitz, and J. B. Sykes, \textit{Electrodynamics of Continuous Media}, 2nd ed. (Elsevier, 1984)

\bibitem{zyablovsky14}A. A. Zyablovsky, A. P. Vinogradov, A. V. Dorofeenko, A. A. Pukhov, and A. A. Lisyansky, ``Causality and phase transitions in PT-symmetric optical systems," Phys. Rev. A \textbf{89}, 033808 (2014).

\bibitem{hagness}S. C. Hagness, R. M. Joseph, and A. Taflove, ``Subpicosecond electrodynamics of distributed Bragg reflector microlasers: Results from finite difference time domain simulations," Radio Sci. \textbf{31}, 931--941 (1996).

\bibitem{boriskina06}S. V. Boriskina, ``Spectrally engineered photonic molecules as optical sensors with enhanced sensitivity: a proposal and numerical analysis," J. Opt. Soc. Am. B \textbf{23}, 1565--1573 (2006).

\bibitem{smotrova13}E. I. Smotrova and A. I. Nosich, ``Optical coupling of an active microdisk to a passive one: effect on the lasing thresholds of the whispering-gallery supermodes.," Opt. Lett. \textbf{38}, 2059--2061 (2013).

\bibitem{smotrova06}E. Smotrova, A. Nosich, T. M. Benson, and P. Sewell, ``Optical coupling of whispering-gallery modes of two identical microdisks and its effect on photonic molecule lasing," IEEE J. Sel. Top. Quantum Electron. \textbf{12}, 78--85 (2006). 

\bibitem{abramowitz}M. Abramowitz and I. A. Stegun, \textit{Handbook of Mathematical Functions} (U.S. Department of Commerce, NIST, 1972). 

\end{thebibliography}
\end{document}